\documentclass[a4paper,12pt]{article}
\usepackage{amsmath}
\usepackage{graphicx}
\usepackage{rotating}
\usepackage{wasysym}
\DeclareGraphicsExtensions{.pdf,.png,.jpg}
\usepackage{ chicago}
\usepackage{ klett}
\usepackage{ def_the}

\usepackage{ Fig_size_A}
\usepackage{ kfig}

\begin{document}
 
 
\title{Free Lunches with Vanishing Risks \\Most Likely Exist}
\author{\\ Eckhard Platen and Kevin Fergusson} 
\date{\today}
\maketitle
\begin{center}
\begin{minipage}[t]{13cm}
 The hypothesis  that there do not exist   free lunches with vanishing risk (FLVRs) in the real market underpins the popular risk-neutral pricing and hedging methodology in quantitative finance. The paper documents the fact that this hypothesis can be safely rejected. It performs extremely accurately the hedging of an extreme-maturity zero-coupon bond (ZCB). This  hedge is part of a portfolio that starts  with zero initial wealth and invests dynamically in a total return stock market index and the savings account  to generate at the maturity date of the extreme-maturity ZCB a strictly positive amount with  strictly positive probability, which represents an FLVR.  The fact that FLVRs naturally exist  in the real  market can be accommodated theoretically under the benchmark approach.
	
\end{minipage}
\end{center}
\vspace*{0.5cm}


\vspace*{0.5cm}
\vspace*{0.5cm}
\newpage
\section{Introduction}
Quantitative methods in finance rely on market models. These models must be mathematically consistent and close to reality to ensure that the  solutions to risk management problems can be reliably applied in practice. For this reason, it has been argued in the literature that it is not realistic to allow in a market model the existence of {\em arbitrage opportunities}. \\ An  arbitrage opportunity is a way of {\em making profits with no initial investment without any possibility of loss}. Arbitrage opportunities may exist briefly in real markets. However, most of the current literature argues that a useful market model must avoid this type of profit; see, e.g., \citeN{Jarrow22}.
 For this reason, the classical mathematical finance theory, as described, e.g., in \citeN{Jarrow22}, relies on its  {\em Fundamental Theorem of Asset Pricing} (FTAP), which was derived in the  seminal work of \citeN{DelbaenSc98}.   The FTAP  has become extremely important as a theoretical underpinning of quantitative methods in finance because it provides the foundations of the widely implemented risk-neutral pricing  methodology. \\

The FTAP employs the notion of a {\em free lunch with vanishing risk} (FLVR),  most generally defined in  \citeN{DelbaenSc98}. It is currently by far the most widely considered  classical notion of arbitrage for continuous-time market models.  The FTAP considers  {\em admissible portfolios}, which are   {\em self-financing}  portfolios that are {\em bounded from below}. Informally said, a market model allows for an FLVR  if there are admissible portfolios that can be chosen arbitrarily close to an arbitrage opportunity. This means, these portfolios start with no wealth, end up with strictly positive wealth (a free lunch) with a probability greater than zero,  and the probability of ending up with negative wealth can be chosen arbitrarily small (the vanishing risk). The  current paper aims to answer the question, {\em do free lunches with vanishing risks most likely exist} in the real market?\\

The philosopher Karl Popper  established a modern set of standards for the scientific methodology; see \citeN{Popper02}. The current paper  follows these standards by aiming to  falsify  the null hypothesis of the FTAP that FLVRs do {\em not} exist. For this purpose, it considers the hedging of extreme-maturity {\em zero-coupon bonds} (ZCBs) that pay approximately one unit of the savings account at maturity. For each extreme-maturity ZCB a  portfolio will be formed that starts  with zero initial wealth and invests dynamically in a total return stock  index and the savings account.  This hedge will be shown to generate very reliably, at the maturity date of the extreme-maturity ZCB, some nonnegative and mostly strictly positive amounts. The empirical study will investigate potential FLVRs for the S\&P500 total return stock market index  and many-extreme maturity ZCBs. The study will falsify in a hypothesis test with  high confidence  the null hypothesis  that FLVRs do not exist.\\

In its second part, the paper reviews parts of the benchmark approach, as described in \citeN{PlatenHe06}. This approach can accommodate  the fact that FLVRs exist in the real  market. By using its benchmark-neutral pricing, as described in \citeN{Platen25a}, and the information-minimizing dynamics of a stock market index, as derived in \citeN{Platen25b},   the theoretical prices and hedging strategies of  ZCBs are derived. These prices and hedging portfolios explain the  presence of  FLVRs in the real market under the benchmark approach.

The paper is organized as follows: Section 2 studies empirically the existence of potential FLVRs and Section 3 explains how FLVRs naturally exist under the benchmark approach.
\section{Potential Free Lunches with Vanishing Risk}
\subsection{Free Lunch with Vanishing Risk}

 Most papers and books currently published in the mathematical finance and quantitative finance literature base their results on the {\em Fundamental Theorem of Asset Pricing} (FTAP); see \citeN{DelbaenSc98}. To illustrate the statement of this important theorem let us
 consider  a continuous-time  financial market model on a filtered probability space $(\Omega,\mathcal{F},\underline{\cal{F}},P)$, satisfying the usual conditions; see, e.g.,  \citeN{KaratzasSh98}.  The filtration $\underline{\cal{F}}$ $=(\mathcal{F}_t)_{t \geq t_0}$ models the evolution of information relevant to the  market model.  
The information available at time $t \in [0,\infty)$ is captured by the sigma-algebra $\mathcal{F}_t$.  The  market model that we consider  describes the dynamics of a {\em savings  account} $S^0_t=1$ and a vector ${\bf {S}}_t$ of $n\in\{1,2,...\}$  {\em risky securities} ${\bf {S}}_t=(S^1_t,...,S^n_t)^\top$ for $t\geq t_0$. For convenience, we discount these securities  by the savings account.  The latter is a locally risk-free portfolio and, in practice, typically approximated by a roll-over short-term zero-coupon bond account.   We  model an idealized   market with continuous trading,    instantaneous investing and borrowing,   short sales with full use of proceeds,    infinitely divisible securities, and      no or small realistic transaction costs. \\

In \citeN{DelbaenSc98} the FTAP has been formulated as follows: \begin{theorem}[Fundamental Theorem of Asset Pricing]
	For an
${\bf {R}}^n$-valued semimartingale  ${\bf {S}}=\{{\bf {S}}_t=(S^1_t,...,S^n_t)^\top, t\geq t_0\}$, the
following are equivalent:\\
1. There exists a probability measure $Q_{S^0}$ equivalent
to $P$ under which ${\bf {S}}$ is a sigma-martingale.\\
2. ${\bf {S}}$ does not permit an FLVR.\end{theorem}

The  precise  definitions of the notions involved in the above formulation of the FTAP  are given in the above-mentioned paper \citeN{DelbaenSc98}. It is not the purpose of the current paper to go  into any details of the formulation of the FTAP or its technically demanding proof.   Instead, it aims to provide  empirical evidence concerning the validity of the key assumption of the FTAP, which is the assumption about the absence of  FLVRs.\\ 
Roughly speaking, the FTAP  states  that the absence of FLVRs in a market model is equivalent to the existence of an equivalent   risk-neutral probability measure $Q_{S^0}$, where the savings account $S^0$ is the num\'eraire and the  equivalent risk-neutral probability measure $Q_{S^0}$ is the pricing measure. This is a crucially important statement for  pricing and hedging in real markets  because the existence of an equivalent risk-neutral probability measure $Q_{S^0}$ provides a  system of unique prices  for the  pricing and hedging of replicable contingent claims. \\

  Risk-neutral pricing is currently the dominant pricing method for the pricing and hedging of derivative contracts; see, e.g., \citeN{Jarrow22}. We define {\em extreme-maturity contracts} as contracts with terms to maturity that reach   at least over a period of $15$ years.  Large volumes of   extreme-maturity contracts exist in the pension and life insurance industry. The costs of these contracts are  important for  millions of  individuals in many countries. Risk-neutral pricing is usually applied in the literature  to extreme-maturity pension and insurance contracts, like variable annuities; see, e.g., \citeN{IFRS17}. Risk-neutral pricing determines these costs  based on the postulate that no FLVRs exist in the real market. Motivated by this fact, the paper raises the following  question:\\

 \noindent{\em \bf Is it possible to falsify in a hypothesis test with high confidence the null hypothesis that FLVRs do not exist in the real  market?}\\
 
To answer this question {\em \bf affirmatively}, it is sufficient to provide clear empirical evidence that a particular type of FLVR exists  with high probability in the real  market. The current paper aims to provide such empirical evidence by pricing and hedging extreme-maturity zero-coupon bonds (ZCBs) less expensively than suggested by  risk-neutral pricing. For a targeted extreme-maturity  ZCB-type  payoff (with a term to maturity of at least $15$ years), it constructs a potential FLVR by  going long in the hedge portfolio that generates  the targeted payoff and short in a  savings account investment  equal to the so-called benchmark-neutral initial price of the ZCB-type payoff that will be derived in Section 3 by following \citeN{Platen25a}.  \\

\subsection{Activity Time}
 
To identify   potential FLVRs in the real stock market
we employ  daily observed   values of a  savings  account-discounted  total return stock index $S_{t_i}$ at the increasing observation times $t_i$ for $i\in\{0,1,...,N\}$, where $N\in\{2,4,6,...\}$ is an   even integer. The savings account-discounted locally risk-free portfolio equals in this denomination the constant $S^0_t=1$ for all $t\geq t_0$.  The discounted stock index $S_t$ represents the risky asset. It is clear from the classical finance theory, as, e.g., described in \citeN{Jarrow22}, that the discounted risk-neutral  price of a ZCB that pays one unit of the savings account at maturity equals the constant $1$. The current paper performs some alternative  pricing and hedging under the benchmark approach and introduces for the discounted stock index its discretely observed so-called {\em activity time} 
\begin{equation}\label{tau}
\tau_{t_i}=\ln\left(\sum_{l=1}^{i}\left(\sqrt{S_{t_{l}}}-\sqrt{S_{t_{l-1}}}\right)^2+e^{\tau_{t_{0}}}\right)
\end{equation}
for $i\in\{1,2,...,N\}$.  It estimates the initial activity time $\tau_{t_0}$ by assuming, at first, a tentative initial activity time value and fitting  a trendline through the resulting discretely observed tentative activity times $\tau_{t_0},\tau_{t_1},...,\tau_{t_\frac{N}{2}}$ via standard linear regression. By varying the tentative initial value $\tau_{t_0}$ of the activity time  in such a way  that the fitted trendline  $\bar \tau_t$ shows the maximum possible $R^2$ value,  it estimates the initial activity time $\tau_{t_0}$. This estimate provides the trendline \begin{equation}
\bar \tau_t=\bar \tau_{t_0}+a(t-t_0)
\end{equation}  with  initial value $\bar \tau_{t_0}$  and  slope $$a=\frac{\bar \tau_{t_{\frac{N}{2}}}-\bar \tau_{t_{0}}}{t_{\frac{N}{2}}-t_{0}}$$ 
for $t\geq t_0$. Note that we fit the initial activity time  to the first half of the available observation period. 
 This means  for the second half of the available observed index values that the calculation of the evolving  trendline will not involve any  information generated in the respective period. \\
 
\begin{figure}
 	\includegraphics[width=\textwidth]{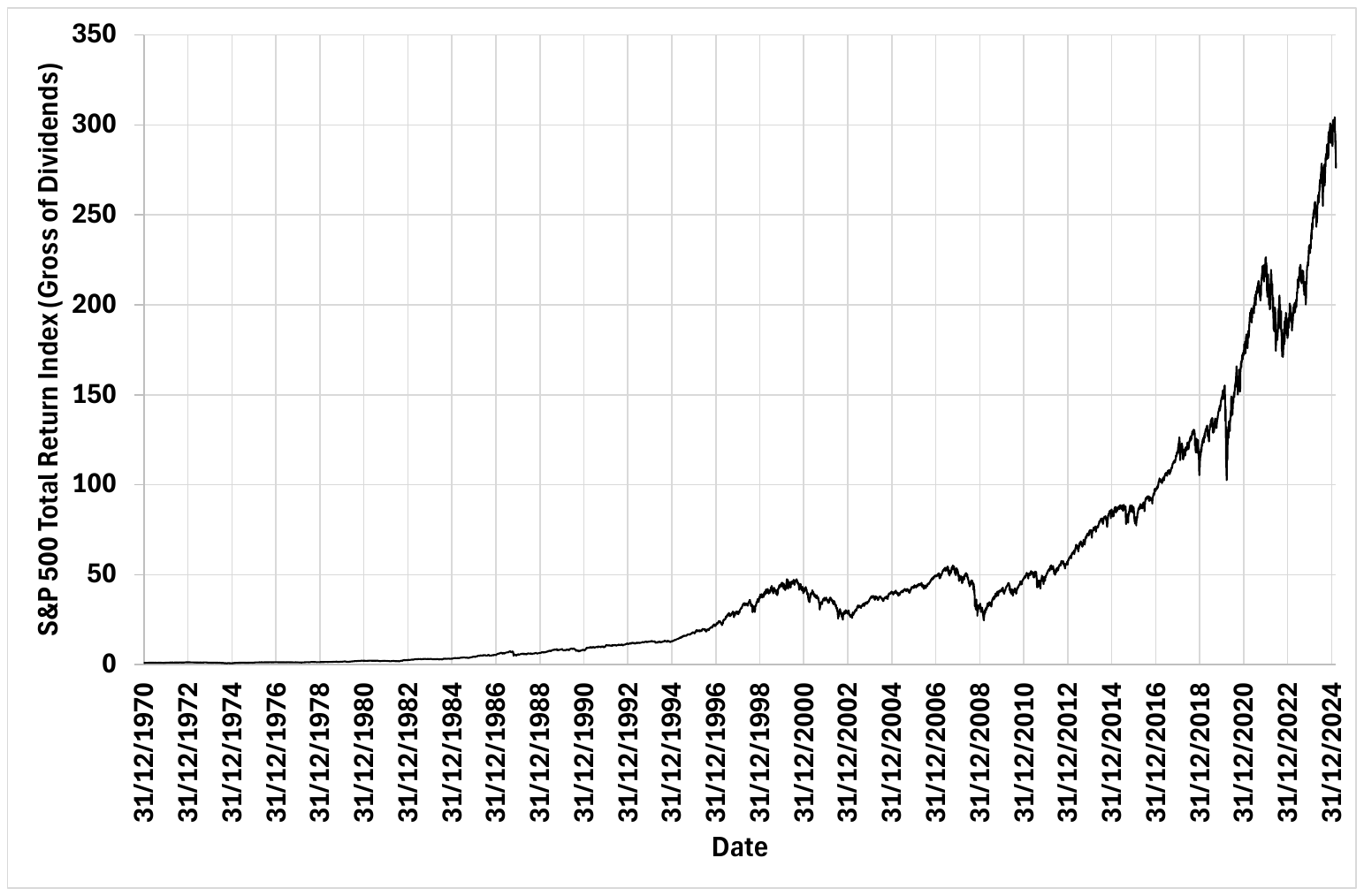}
 	\caption{\label{Figure1}Discounted S\&P500 Total Return Stock Index $S_t$ over the period from 31st December 1970 to 11th March 2025.}
 \end{figure}

\begin{figure}
 	\includegraphics[width=\textwidth]{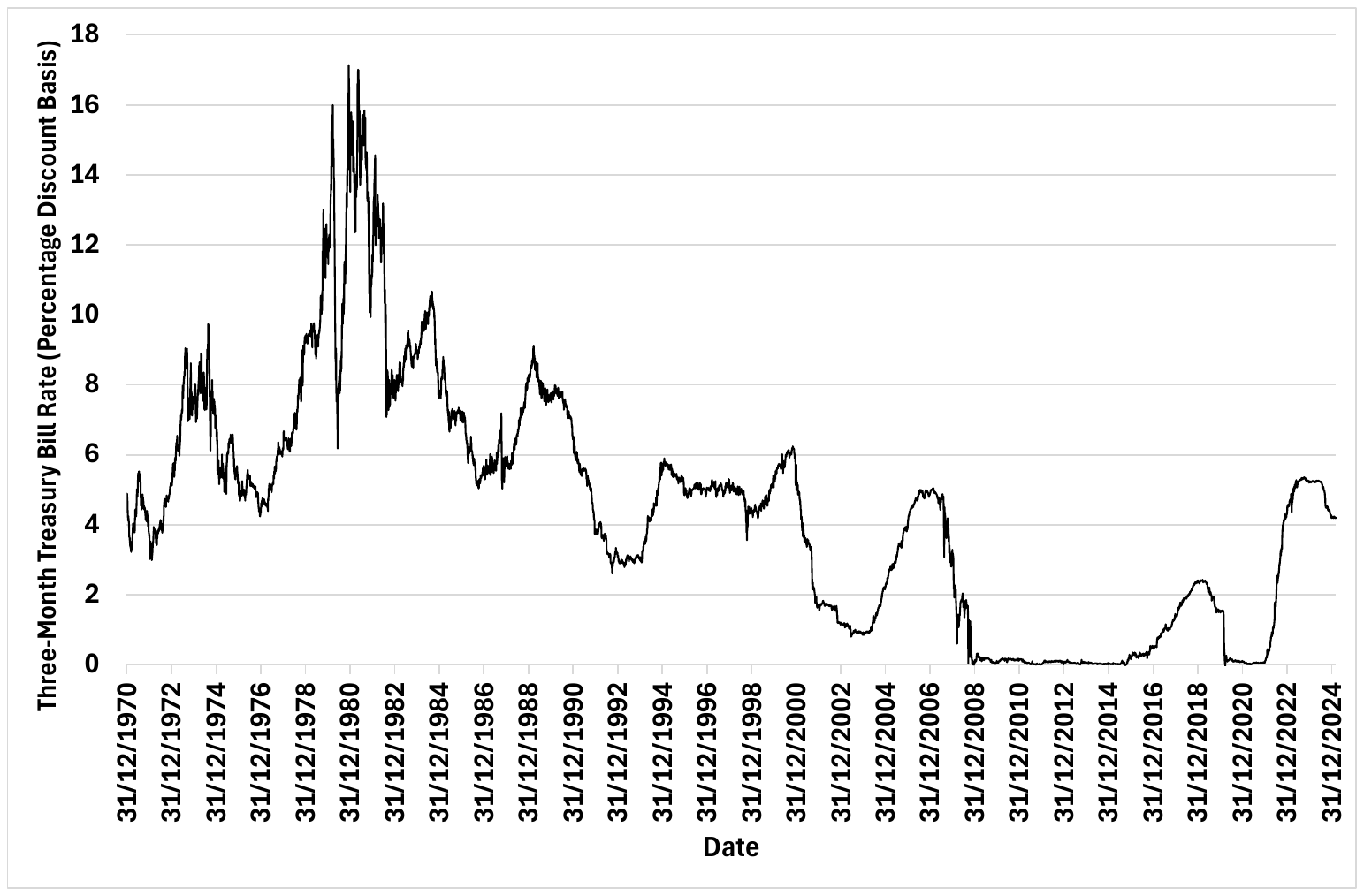}
 	\caption{\label{Figure2}Three-month US Treasury bill rates over the period from 31st December 1970 to 11th March 2025.}
 \end{figure}

 To illustrate the above-described estimation and calculation of the evolving activity time, we employ the daily observed savings account-discounted S\&P500. Daily S\&P500 Total Return Index data over the period from 31st December 1970 to 11th March 2025 is sourced from \citeN{SP500} and  depicted in Figure~\ref{Figure1}.
 The savings account is approximated by a roll-over  three-month US Treasury Bill account.   
 Respective daily US interest rate data over the period $t_0=$31 December 1970 to 11 March 2025 is sourced from \citeN{DTB3} and  depicted in Figure~\ref{Figure2}.
 Because the interest rates are quoted on a discount basis, the daily series of US dollar-denominated savings account values $\bar S^0_{t_0},S^0_{t_1},...$ is constructed recursively via
 $$
 \bar S^0_{t_{i+1}} = \bar S^0_{t_{i}} \times \left( 1- \frac{r^{DTB3}_{t_{i+1}}}{100}\times \frac{90-d(t_i,t_{i+1})}{360}\right) / \left( 1- \frac{r^{DTB3}_{t}}{100}\times \frac{90}{360}\right),
 $$
 with initial value $\bar S^0_{t_0}=1$, where $r^{DTB3}_{t}$ denotes the discount rate of 3-month treasury bills in the secondary market and $d(t_i,t_{i+1})$ the number of days from time $t_i$ to time $t_{i+1}$.
 
\begin{figure}
 	\includegraphics[width=\textwidth]{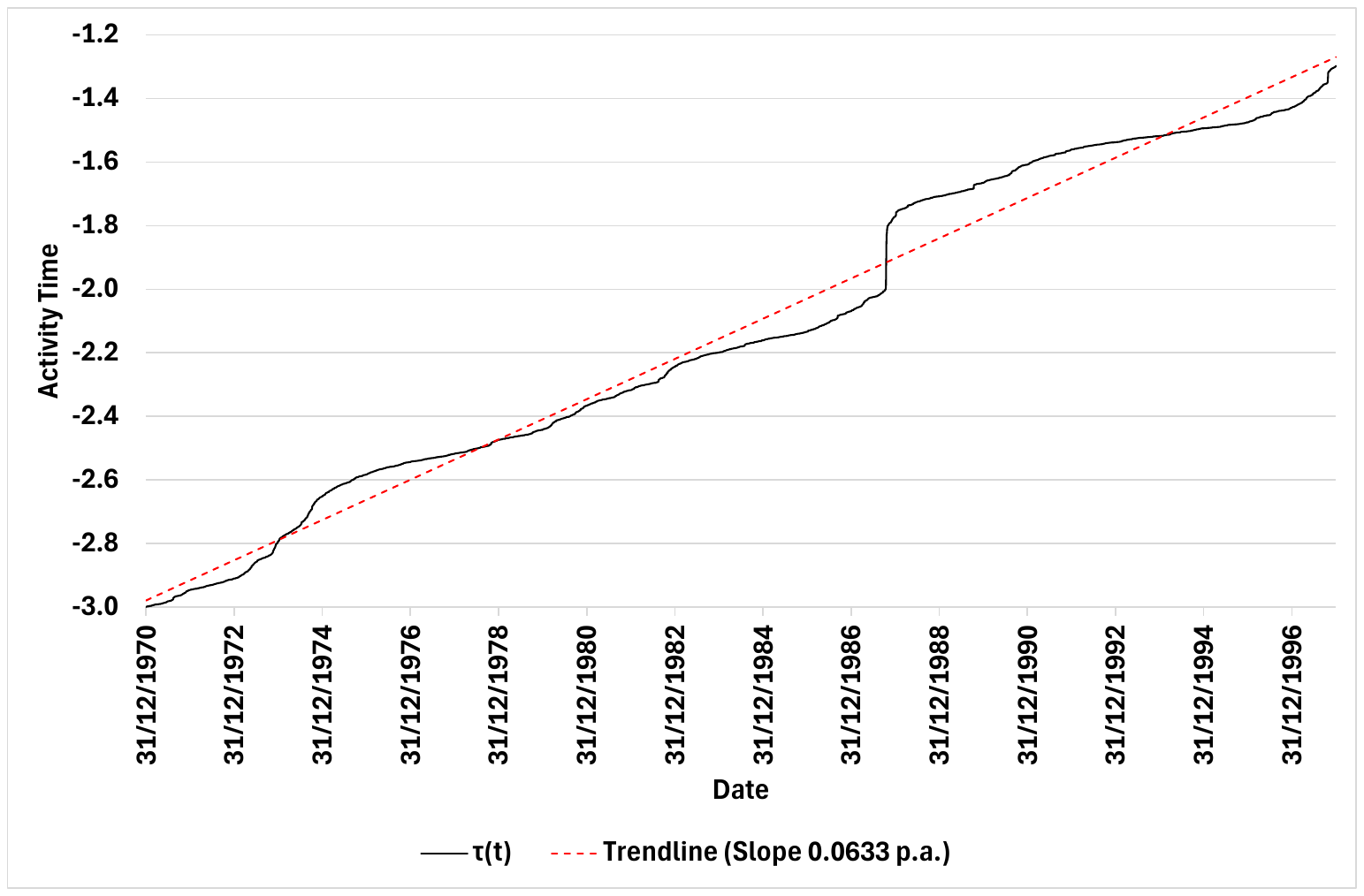}
 	\caption{\label{Figure3}Activity time $\tau_t$ with trendline calculated  from  savings account-discounted S\&P500 data for the period from 31st December 1970 to 31st December 1997.}
 \end{figure}

 In Figure \ref{Figure3} the calculated activity time $\tau_t$ is displayed for the  savings account-discounted S\&P500  when using the data from $t_0=$ 31 December 1970 until $t_{\frac{N}{2}}=$ 31 December 1997. One notes that the activity time shows some randomness
 but evolves on average  linearly. The resulting trendline, which is formed from the S\&P500 observations of the first half of the available observation period, covering the period from 31st December 1970 to 31st December 1997, is  displayed in Figure \ref{Figure3}. The $R^2$ value for the respective linear regression amounts to $98.01\%$, which supports the underlying assumption that the activity time evolves approximately linearly. The following calculations of the ZCB prices for periods in the second half of the available observation period will employ the extension of the trendline of the activity time for the second half of the observation period. The values of the trendline do not rely on any observations of the S\&P500 from that period.
\subsection{Potential FLVR of a Zero-Coupon Bond}
 According to the definition of an FLVR  in \citeN{DelbaenSc98}, an example of a potential FLVR would be given by a discretely observed portfolio process $V=\{V_{t_i},i\in\{\underline i,...,\bar i\}\}$ with $\underline i\in\{\frac{N}{2},...,N-1\}$  that starts at some observation time $t_{\underline i}$  with value  \begin{equation}V_{t_{\underline i}}=0 \end{equation} and ends up at some observation time $t_i$ with $i\in\{\underline i+1,...,N\}$ with some positive value $V_{t_i}>0$ with a probability
\begin{equation}P(V_{t_i}>0)>0\end{equation} that is  greater than zero, where the portfolio process $V$ remains always bounded from below.\\

  Roughly speaking, when forming a potential FLVR we go   long in a hedge that replicates the payoff of a ZCB that pays one unit of the savings account at maturity and we go short in the initial price of the ZCB invested in the  savings account. Since the activity time is random and hedging occurs, in reality, at discrete times,  we  consider instead of a  ZCB that pays at maturity one unit of the savings account,  an {\em approximate zero-coupon bond} (AZCB) in savings account-denomination with maturity $t_{\bar i}$  and nonnegative {\em ZCB-type
  payoff}
  \begin{equation}\label{HT} H_{t_{\bar i}}=1-\exp\left\{-\frac{S_{t_{ i}}}{2\max(e^{\bar\tau_{t_{\bar i}}}-e^{\tau_{t_{ \bar i}}},0)}\right\},\end{equation} which approaches asymptotically $1$ when the activity time step size tends to zero. More precisely, the second term on the right-hand side of the above equation equals $0$ when  $\tau_{t_{\bar i}}\ge \bar \tau_{t_{\bar i}}$, and it vanishes asymptotically for $\tau_{t_{\bar i}}\rightarrow \bar \tau_{t_{\bar i}}$ so that  the  ZCB-type
  payoff always equals asymptotically $1$. In the case  $\tau_{t_{\bar i}}<\bar \tau_{t_{\bar i}}$, the AZCB payoff is in our study slightly less than $1$ and strictly positive. Its difference from $1$ turns out to be  extremely small for the extreme-maturity AZCBs we will consider.\\
  
  We determine the AZCB price $ P(t_{\underline i},t_{\bar i})$ at the  initiation of the hedge at the time $t_{\underline i}$ by using the formula
   \begin{equation}\label{barP} P(t_{\underline i},t_{\bar i})=1-\exp\left\{-\frac{S_{t_{ \underline i}}}{2\max(e^{\bar\tau_{t_{\bar i}}}-e^{\tau_{t_{\underline i}}},0)}\right\}\end{equation} 
    for $i \in\{ \underline i,\bar i-1\} $ and $\bar i \in\{\underline i+1,...,N\}$. This AZCB pricing formula has been suggested in \citeN{Platen25a}  for an illustration of the benchmark-neutral pricing methodology, which  will be summarized in Section 3. \\
   
    To detect a potential FLVR the  AZCB   price process is approximated by a self-financing hedge portfolio $ Z(t_{ i},t_{\bar i})$ with initial value \begin{equation} Z(t_{ \underline i},t_{\bar i})= P(t_{\underline i},t_{\bar i})\end{equation} and  value
\begin{equation}
 Z(t_{ i},t_{\bar i})= Z(t_{  i-1},t_{\bar i})\left(1+\pi_{t_{i-1}}\left(\frac{S_{t_{i}}}{S_{t_{i-1}}}-1\right)\right)  
\end{equation}
for $  Z(t_{  i-1},t_{\bar i})<1$ and  value $Z(t_{ i},t_{\bar i})= Z(t_{  i-1},t_{\bar i})$ otherwise. Here
\begin{equation}\label{pi}
\pi_{t_{i-1}}=\left( 1 - \frac{1}{ Z(t_{  i-1},t_{\bar i}))} \right)\ln ( 1- Z(t_{  i-1},t_{\bar i}))
\end{equation} denotes the fraction invested in the index value $S_{t_{i-1}}$ for $i \in\{ \underline i+1,\bar i\}$.\\
The self-financing portfolio $Z(t_{  i},t_{\bar i})$ allows  forming   the {\em potential FLVR}
\begin{equation}
V(t_i)= Z(t_{ i},t_{\bar i})-  P(t_{ \underline i},t_{\bar i})
\end{equation}
at the observation time $t_i$ for  $i \in\{ \underline i,\bar i\}$ as the difference between the value of the hedge portfolio and the initial value of the AZCB.
 If the hedging of the AZCB were continuous, then the above potential FLVR would be the difference between the respective strictly positive AZCB price and its  bounded initial value. This means that the value of $V$ would be theoretically bounded from below by the lower bound $-1$ if the hedge were continuous. Since the hedge errors from discrete-time hedging will be shown to be extremely small, this fact will allow us to interpret the employed portfolio $V$ as being bounded from below and, therefore, as an admissible portfolio  in the sense of the definition of an FLVR and the  FTAP.\\

\subsection{Potential FLVRs for AZCBs}
\label{sec:potentialflvrs}
For illustration, we continue employing the S\&P500 data and consider the pricing and hedging of the AZCB that is initiated in the middle  of the observation period on $t_{\underline i}=$ 31st December 1997 and matures on $t_{\bar i}=t_N=$ 11th March 2025 at the end of the dataset. Its term to maturity is greater than $20$ years, which means we can interpret the AZCB as an extreme maturity contract.\\
\begin{figure}
	\includegraphics[width=\textwidth]{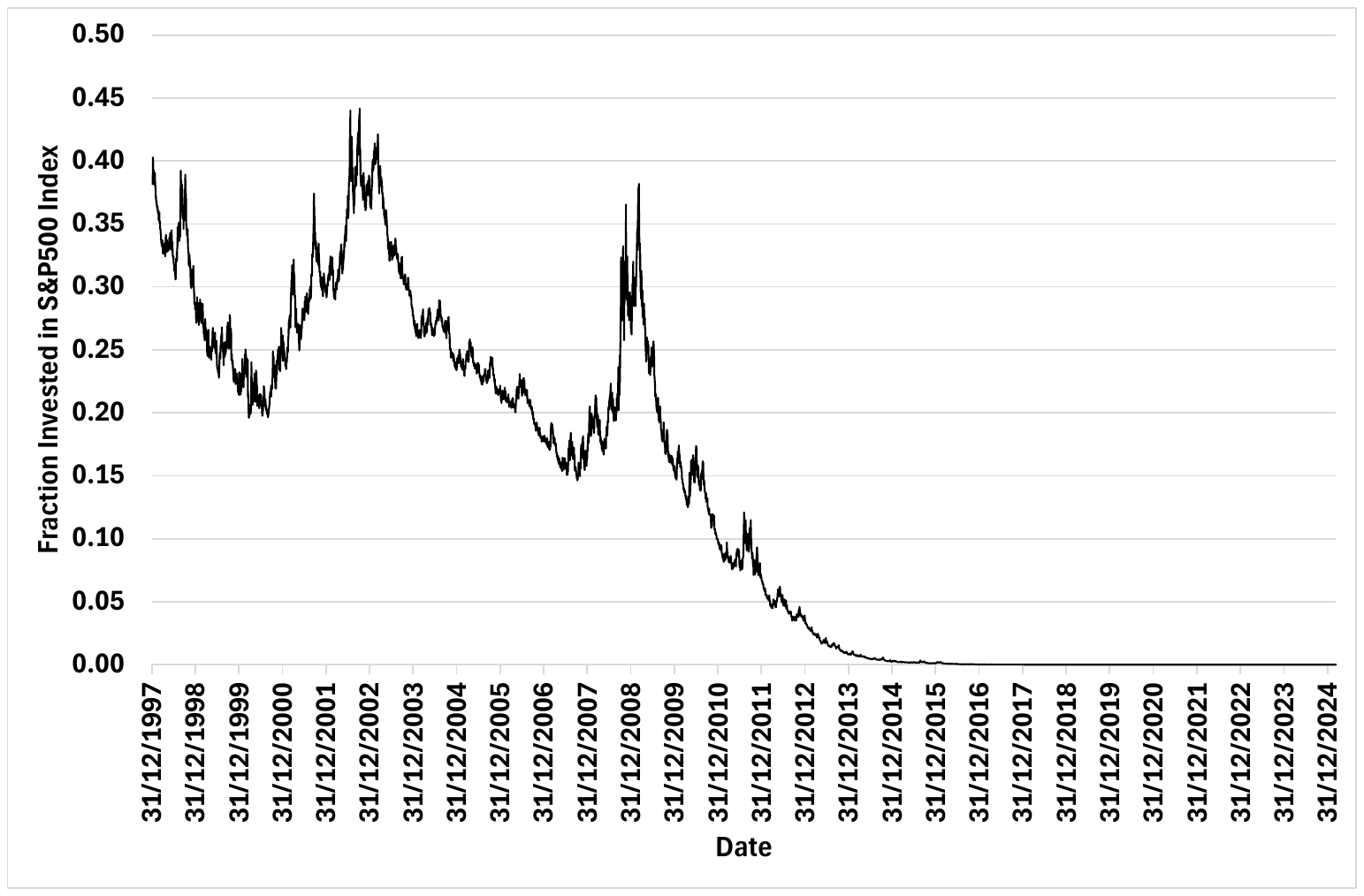}
	\caption{\label{Figure4}Fraction $ \pi_t$ of the AZCB invested in the S\&P500 with maturity on 11 March 2025.}
\end{figure}
In Figure~\ref{Figure4} we show the fraction that the hedge portfolio invests in the S\&P500. One notices  at the beginning that the weight for the investment in the index is relatively large (about  $40\%$). It  becomes, on average, smaller  when approaching maturity, where it finally vanishes.
\begin{figure}
	\includegraphics[width=\textwidth]{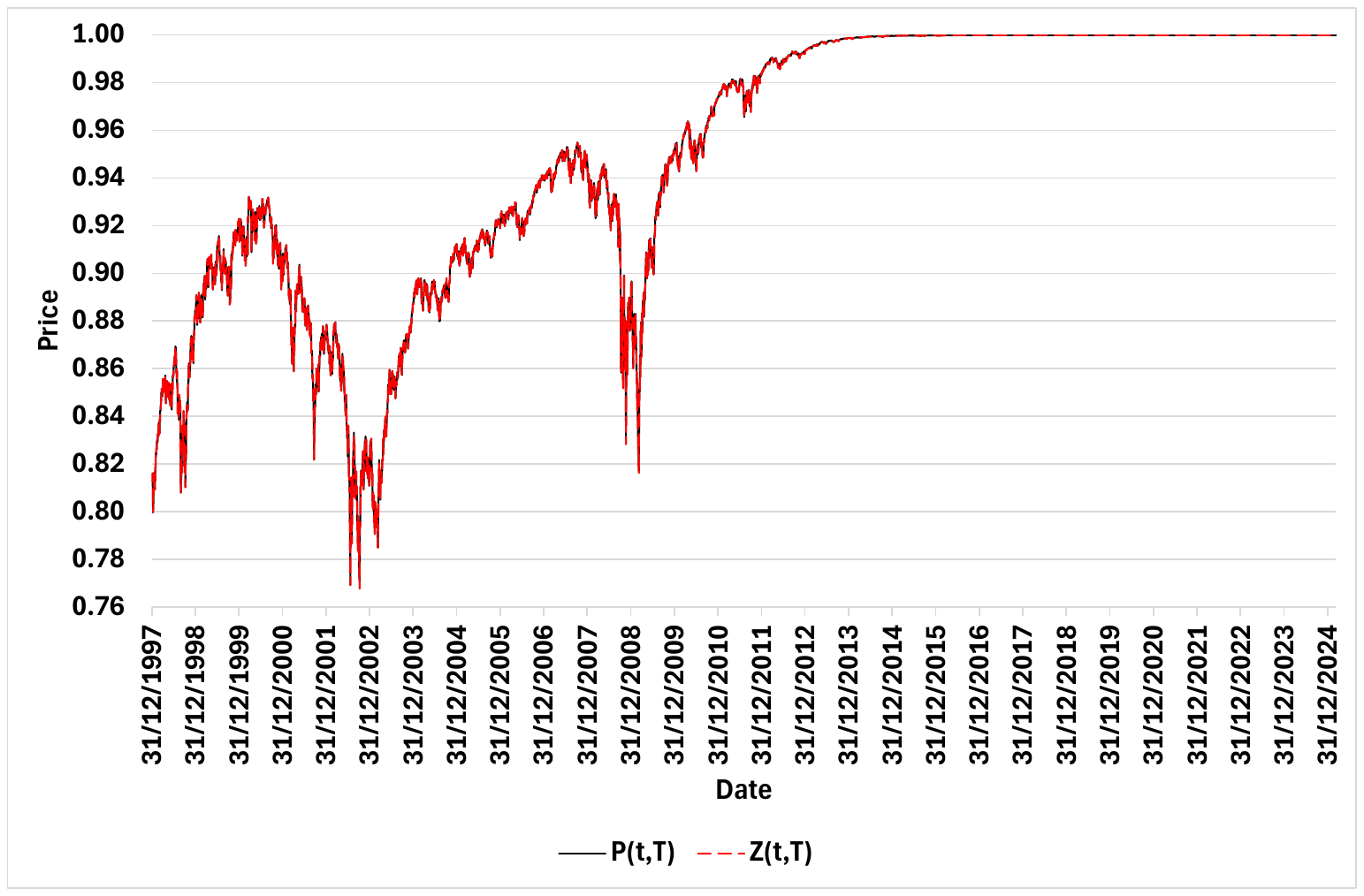}
	\caption{\label{Figure5} Theoretical AZCB price $P(t,T)$ and approximating hedge portfolio $Z(t,T)$.}
\end{figure}
We plot in Figure~\ref{Figure5} the trajectory of the  theoretical price
\begin{equation}
 P(t_i,t_{\bar i})=1-\exp\left\{-\frac{S_{t_i}}{2\max(e^{\bar\tau_{t_{\bar i}}}-e^{\tau_{t_i}},0)}\right\}
\end{equation}
of the AZCB together with its  approximating discrete-time hedge portfolio $Z(t_{ i},t_{\bar i})$ for $i\in[\underline i,\bar i]$. One notes that their difference is so small that it is almost not visible in this figure  even though the hedge was performed over a period of more than $20$ years. To make their difference visible, the hedge error \begin{equation}C_{t_i}= P(t_i,t_{\bar i})-Z(t_i,t_{\bar i})\end{equation} is displayed in Figure~\ref{Figure6}, where we notice that the absolute value of the hedge error remains less than $0.0006$ for a contract that targets at maturity the value $1.0$. This astonishing accuracy of the hedge appears to be a consequence of employing for pricing the information-minimizing market dynamics of the stock index, which became revealed under the benchmark approach in \citeN{Platen25b}.\\

\begin{figure}
	\includegraphics[width=\textwidth]{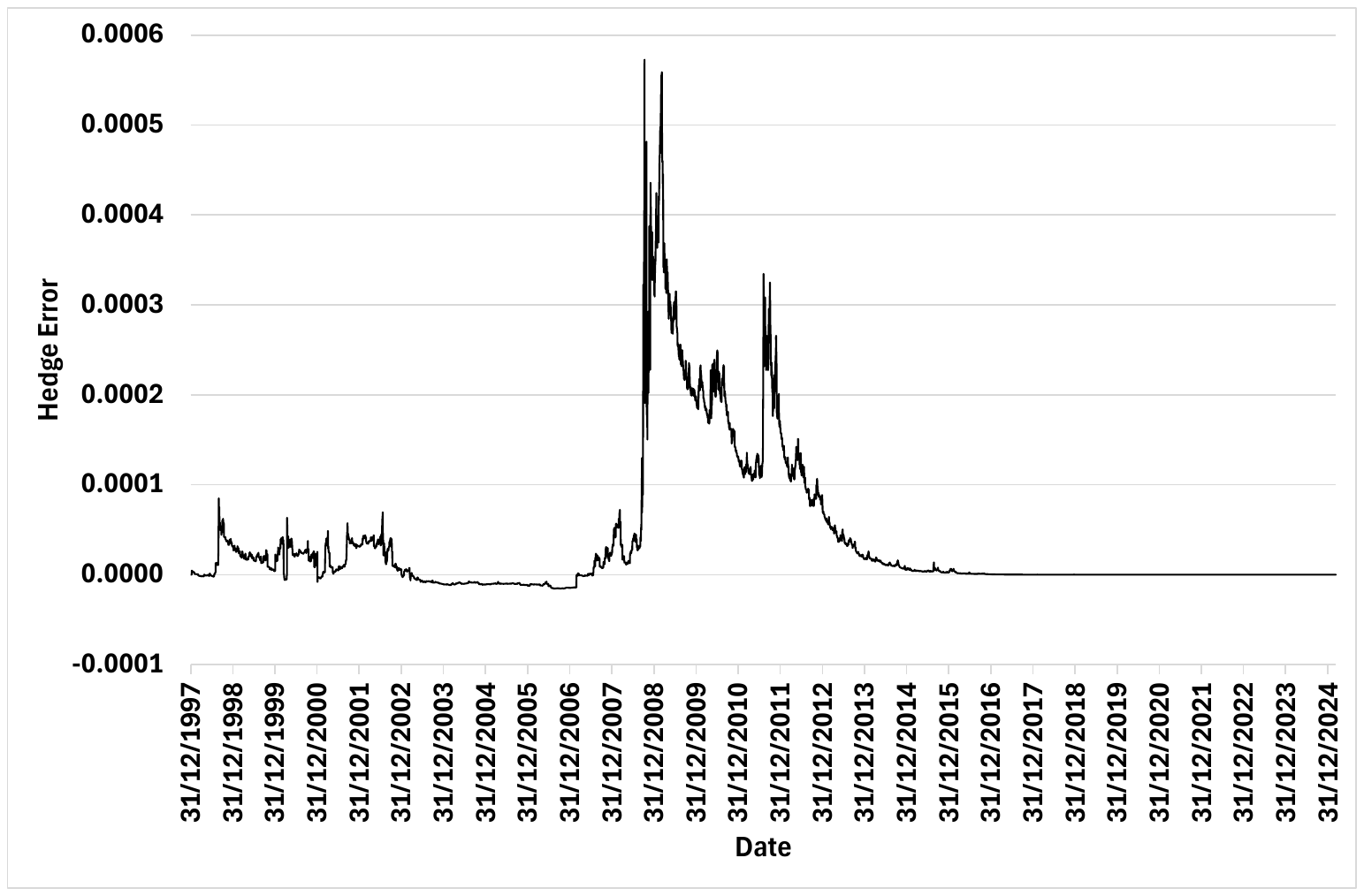}
	\caption{\label{Figure6} Hedge error $C_{t_i}$ for AZCB.}
\end{figure}
\begin{figure}
\includegraphics[width=\textwidth]{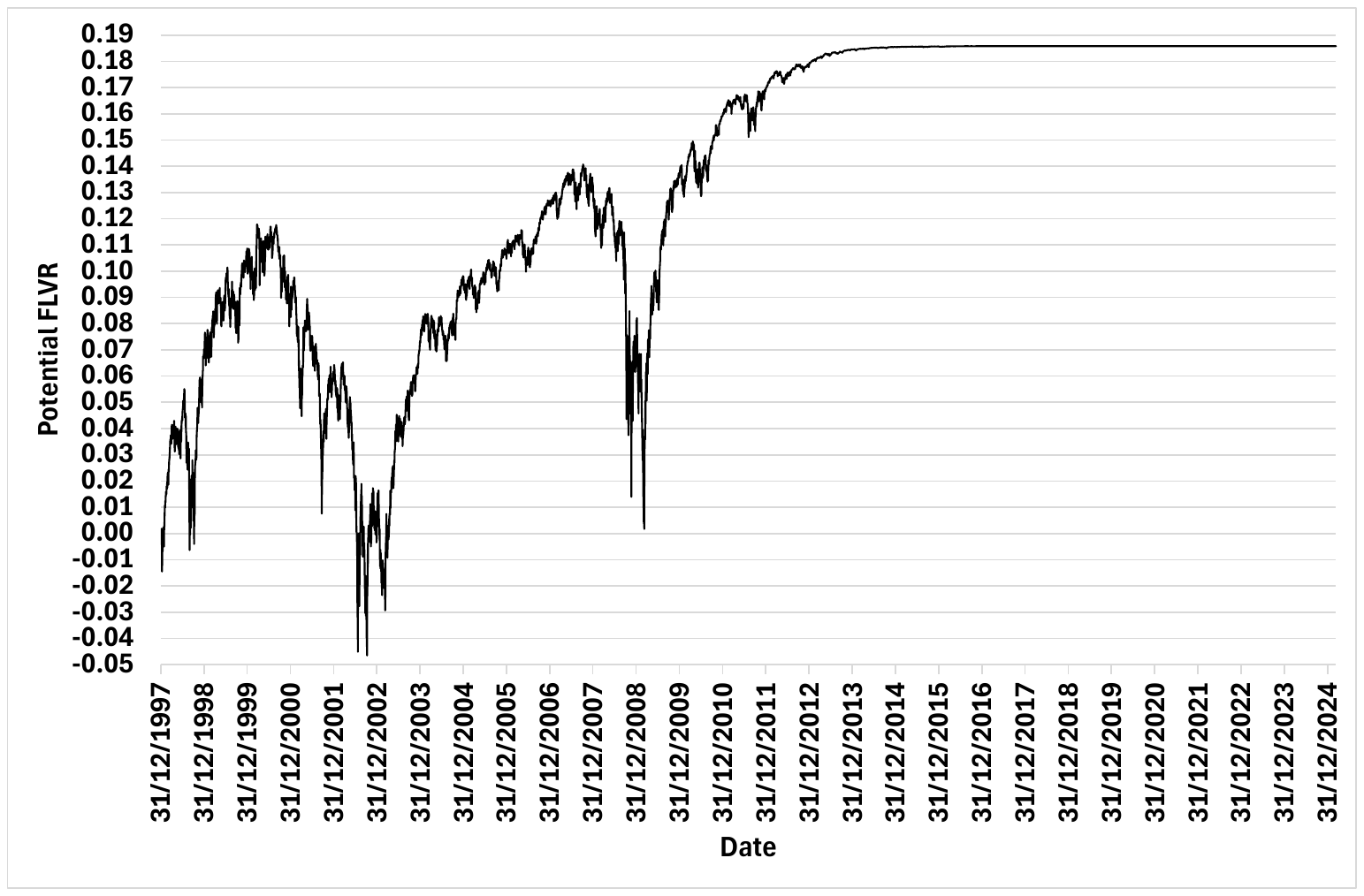}
\caption{\label{Figure7} Potential FLVR $V(t_i)$.}
\end{figure}

 Larger  absolute hedge errors occurred during and  after the Great Financial Crisis (GFC), where the activity time evolved initially much faster than in other periods, which most likely caused some discrete-time hedge errors.  Discretization errors   result from discrete-time hedging instead of  continuous hedging similarly as for discrete-time numerical methods when solving numerically stochastic differential equations; see \citeN{KloedenPl92}.  After the GFC the absolute hedge errors declined visibly. This indicates that these hedge errors are probably mostly caused by  deviations from the in \citeN{Platen25b} derived information-minimizing market dynamics, which became disrupted by the GFC. The observed {\em maximum absolute hedge error} amounts to \begin{equation}\bar C=\max_{i\in \{\underline i,...,\bar i\}}|C_{t_i}|= 0.000573 \end{equation} We plot the respective potential FLVR $V_{t_i}$ for $i\in[\underline i,\bar i]$ in Figure~\ref{Figure7} and notice that its value at maturity is clearly above $0.18$. 
Since this value is substantially greater than zero and the maximum absolute hedge error turns out to be by several magnitudes smaller, this hedge indicates that the observed potential FLVR is very likely an FLVR. \\

\subsection{Testing the Null Hypothesis}

One may argue that the above potential FLVR occurred accidentally. However, this would not be consistent with the observed extremely small hedge error, which means the extremely accurate replication of the AZCB payoff by the approximating discrete-time hedge portfolio. 

To solidify the empirical evidence in this direction, we follow Popper's modern scientific methodology, as described in \citeN{Popper02}, and aim at falsifying the null hypothesis of the FTAP that FLVRs do not exist by repeating the hedge experiment for other extreme-maturity AZCBs.
In this context, it is worth listing the four main aspects of Popper's modern scientific methodology:\\

\noindent	(i) Role of criticism: The scientific method to improve  human knowledge should be characterized by critical examination and debate, where scientists actively try to find flaws in existing theories.\\
	
\noindent(ii) Falsifiable hypothesis:
A scientific theory  is considered falsifiable if it can be logically contradicted by an empirical test.\\

\noindent(iii) The null hypothesis:
 Popper's philosophy of falsification is often used as a basis for testing a null hypothesis, which is a statement based on observable variables, where the goal is to find evidence against it. \\
 
\noindent(iv) Rejecting versus failing to reject:
In hypothesis testing, one either rejects the null hypothesis (meaning there is evidence against it) or fails to reject it (meaning there is not enough evidence to conclude that the null hypothesis is false). \\

\begin{figure}
	\includegraphics[width=\textwidth]{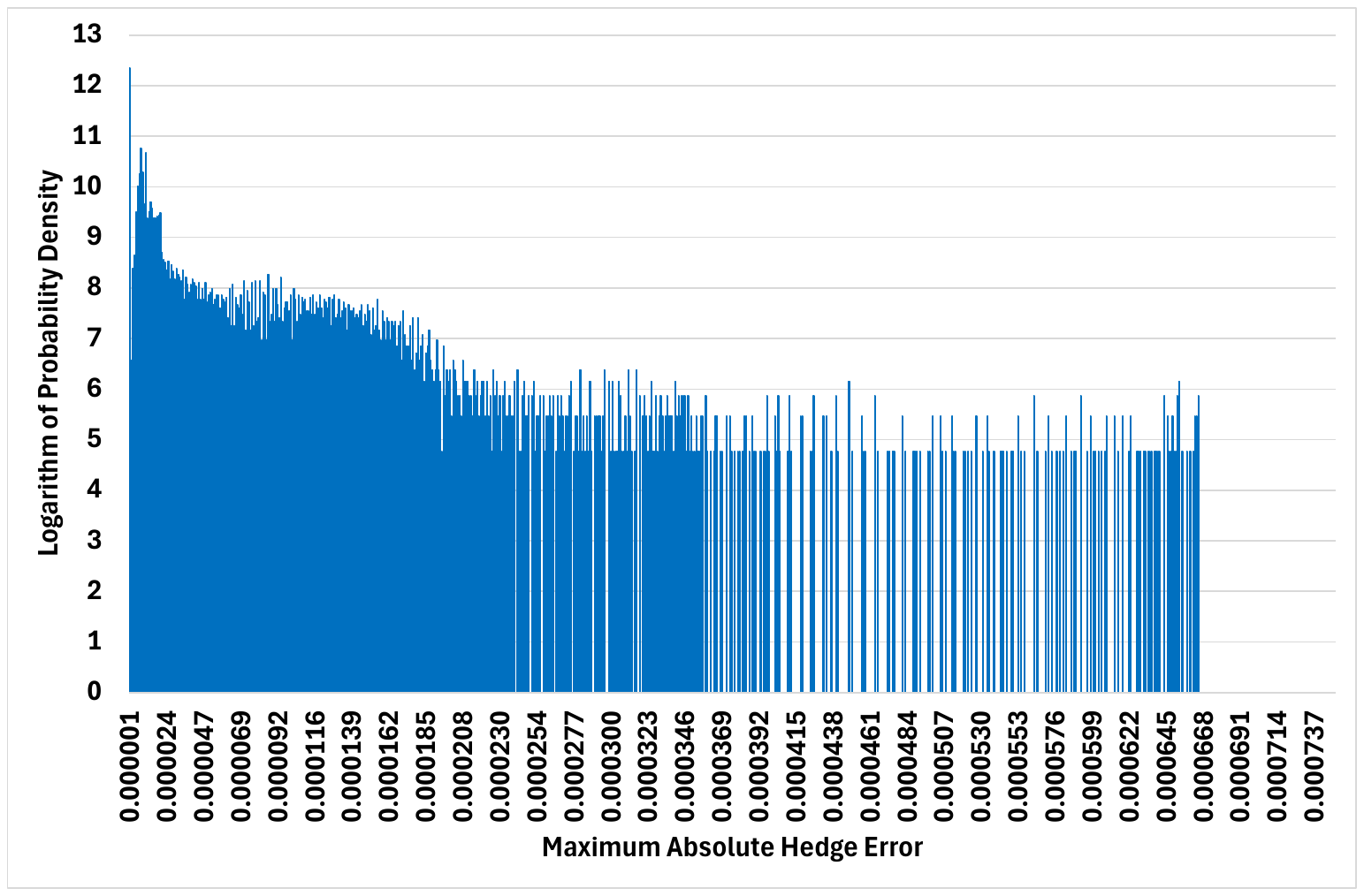}
	\caption{\label{Figure8} Histogram of maximum absolute hedge errors.}
\end{figure}

\begin{figure}
	\includegraphics[width=\textwidth]{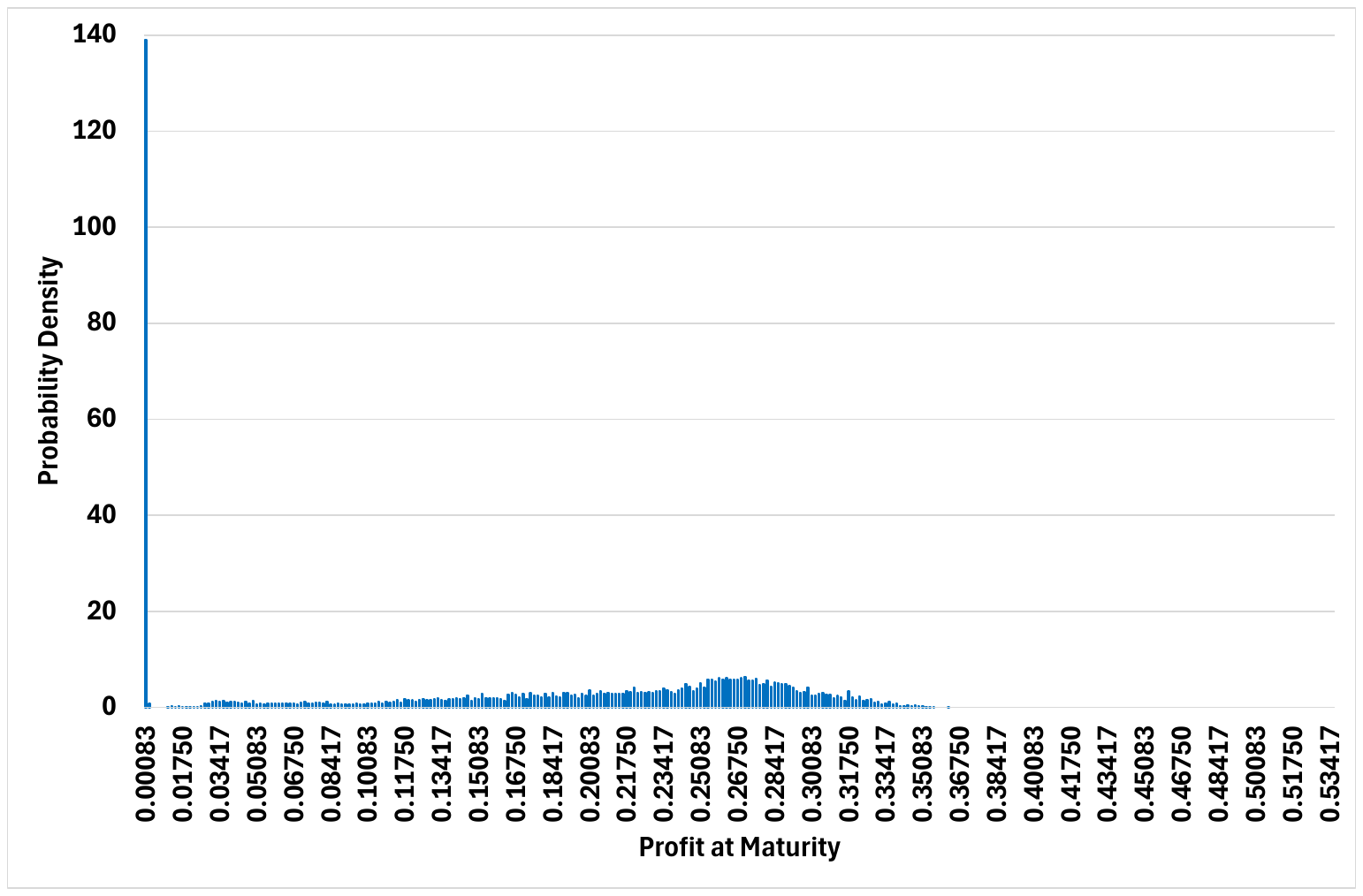}
	\caption{\label{Figure9} Histogram of potential FLVRs, the  profits at maturity.}
\end{figure}

Regarding the first aspect (i) of Popper's modern scientific methodology, the current paper aims to stimulate a critical examination and debate about the key assumption of classical mathematical finance theory, which has been highly relevant to  quantitative finance.\\
By utilizing the second aspect (ii), the current paper repeats the above empirical experiment for many other extreme-maturity AZCBs.  It  selects for the S\&P500 data set all beginnings of a month  that represent possible initiation times for extreme-maturity AZCBs  with monthly set terms to maturity from $15$  to $17$ years. 
For the resulting large set of $n=8475$ extreme-maturity AZCBs it performs the above-described calculation of the  potential FLVRs and monitors the respective maximum absolute hedge errors. The histogram of the maximum absolute hedge errors for the  AZCBs studied is displayed in Figure~\ref{Figure8}.   One notes that all maximum absolute hedge errors remained below {\bf 0.0007}. The standard deviation  of the maximum absolute hedge errors remains below {\bf 0.0001}. Recall that this is the hedge error where the targeted payoff is about  {\bf 1.0}. The histogram for the observed potential FLVRs is shown in Figure ~\ref{Figure9}. All potential FLVRs are  strictly positive. 
  The maximum absolute hedge errors turn out to be by magnitudes smaller than the  mean {\bf 0.1680} of the obtained potential FLVRs. For illustration, we display in Figure~\ref{Figure9} the histogram of the potential FLVRs, which can be interpreted as the  profits at maturity. \\ 
Regarding the third aspect (iii), it is not so easy to formulate a null hypothesis that takes into account the fact that the potential FLVRs  vary. Most important is the empirical fact that the observed potential FLVRs are strictly positive. This is due to the fact that their theoretical values are strictly positive and their hedge errors are extremely small.\\ By aiming to keep  in line with popular  statistical practice, let $\mu$ denote the true mean of the potential FLVRs.  The null hypothesis could be stated as\\ $$ H_0: \mu = 0,$$
 with the corresponding alternative hypothesis:
 $$ H_1: \mu >0.$$
 
 This leads us to the fourth aspect (iv), which  answers our original question by obtaining the outcome of a respective null hypothesis test with the following details:\\ 

\noindent For each of the $n$ repetitions of the hedge experiment conducted in Section~\ref{sec:potentialflvrs},  the random variable $V_k$, $k\in\{1,2,\ldots ,n\}$, denotes the   $k$-th potential FLVR value at the respective AZCB's maturity date. To interpret the findings as is common in statistical practice, we treat for our test the outcomes of the experiments  as if each $V_k$ were independent and identically distributed. Furthermore, let
$m_V = \sum_{k=1}^n V_k / n$ be the sample mean and $s_V^2 =  \sum_{k=1}^n (V_k - m_V)^2 / (n-1)$
be the sample variance.  Since $m_V$ is an unbiased estimate of $\mu$, an \\ \vspace{0.5cm} intuitive decision rule is given by:\\ \vspace{0.5cm}
 Reject $H_0$ in favor of $H_1$ if $m_V$ is much larger than $0$.\\
 The distribution of the sample mean $m_V$ is difficult to determine exactly. It is common statistical practice, as described, e.g., in \citeN{Hogg2019} or Section 2.2 in ~\citeN{PlatenHe06}, to treat
under $H_0$ the test statistic 
$$T = \frac{m_V }{ s_V/\sqrt{n}}$$ 
 as it would approximately follow a Student-t distribution with $n-1$ degrees of freedom.
The summary statistics from the $n=8475$ observations of potential FLVRs  are $\hat{m}_V = {\bf 0.1680}$ and $\hat{s}_V = {\bf 0.1135}$. 
We employ  $\alpha =$ {\bf 0.000001} as the level of significance of our statistical test and denote by $t(1-\alpha,n-1) = F_{t(n-1)}^{-1} (1-\alpha)=4.757$  the inverse function of the Student-t distribution $F_{t(n-1)}(.)$ with $n-1=8474$ degrees of freedom.
 Using this, we obtain a test\\ \vspace{0.5cm}with the level of significance $\alpha$, where the decision rule is given by: \\ \vspace{0.5cm}Reject $H_0$ in favor of $H_1$ if \hspace{1.0cm} $T \geq t(1-\alpha,n-1)$. \\
This means, we reject $H_0$ if \hspace{1.0cm}
$	\hat m_V\geq t(1-\alpha,n-1) \frac{\hat s_V}{\sqrt{n}} = {\bf 0.0059}.$ \vspace{0.5cm}\\
Our test statistic $\hat m_V=$ {\bf 0.1680} lies by magnitudes above {\bf 0.0059}, so that $H_0$ is clearly rejected; 
see, e.g., Page 269 of \citeN{Hogg2019}. 
The level of significance of our statistical test  equals  the probability of incorrectly rejecting $H_0$, which equals $\alpha=0.000001$ and is extremely low. \\

\noindent In summary, the outcome is as follows:\\

\noindent{\em \bf  It is possible to falsify in a hypothesis test  the null hypothesis that FLVRs do not exist in the real  market with extremely low probability of incorrectly rejecting it.}\\


For the hedging of derivatives, the transaction costs are typically considered  to be negligible in the literature. However,  one may ask the following question: How much does the inclusion of transaction costs   remove from the mean of the observed potential FLVRs? To clarify this question, the above study has been repeated by including different levels of proportional transaction costs.  Of course, as should be expected, the resulting mean of the potential FLVRs becomes smaller. However, it remains still, on average, above $75\%$ of the previously detected potential FLVRs for a more than realistic level of about $50bp$ proportional transaction costs. This leads  clearly still to the rejection of the null hypothesis.\\ It is worth mentioning that a large market participant who would dynamically hedge an extreme-maturity ZCB in the way as described, would buy the stock index when it declines and would sell the stock index when it  increases. This would provide    welcome liquidity in the stock market, in particular, at major downward  moves. Brokers typically reward such a liquidity provider with extremely low or even zero transaction costs. \\

The  hedges of the  extreme AZCBs constructed for the S\&P500 turn out to be extremely accurate even over several decades of hedging. Analogous outcomes were obtained by the authors for  total return stock indices of the stock markets of the G7 economies. This means that FLVRs cannot be lightly assumed to be absent in developed stock markets as is often postulated in most of the quantitative finance literature.  As is documented by the above study, as well as in \citeN{BaroneadesiPlSa24} and \citeN{FergussonPl23}, the putative risk-neutral pricing measure is   providing   for many extreme-maturity contracts prices  that are more expensive than necessary. This indicates a need to make finance theory and its quantitative methods  consistent with the above-stated fact that the absence of FLVRs cannot be realistically assumed.\\ 

\section{Benchmark-Neutral Pricing}
\subsection{Stock GOP and Extended-Market GOP}
Given the above findings, it is of interest to provide a theoretical understanding of the fact that the null hypothesis on the absence of FLVRs is strongly rejected by  market data.  The following summarizes    several results of the benchmark approach, described in \citeN{FilipovicPl09}, \citeN{PlatenHe06},  \citeN{Platen25a}, and \citeN{Platen25b}, which offer a theoretical explanation of the above-documented stylized empirical fact. \\

We continue  denoting by $S^0_t$  the {\em savings account}, which equals the constant $S^0_t=1$ for all $t\geq t_0$ and  denominate all securities in  terms of the savings account. By Theorem 3.1 in \citeN{FilipovicPl09},
the dynamics of the {\em stock GOP} $S_t$, which is the {\em growth optimal portfolio} (GOP) of the stocks (without the savings account in the respective investment universe), are in a continuous market of stocks characterized by the SDE
\begin{equation}
\label{Eqn2.1}
\frac{dS_t}{S_t} = \lambda^*_t dt + \theta_t (\theta_t dt + dW_t),
\end{equation}
for $t\geq t_0$ with $S_0>0$, where $\lambda_t^*$ is the so-called generalized risk-adjusted return and $\theta_t$ the stock GOP  volatility and so-called generalized market price of risk for the  Brownian motion $W_t$.  As the notation indicates, we interpret the stock market index   $S_t$ as the  GOP of the stocks.\\  To price and hedge contingent claims, we consider a market formed by the savings account $S^0_t=1$ and the stock GOP $S_t$. In this market we form the GOP $S^{**}_t$, which is the GOP of the market consisting of $S^0_t=1$ and  $S_t$ and we call $S^{**}_t$ the {\em extended-market GOP}.  By  employing Equation 2.2 of \citeN{Platen25a} and Lemma 7.1 of \citeN{FilipovicPl09}
the extended-market GOP satisfies the SDE
$$
\frac{dS^{**}_t}{S^{**}_t} =  \sigma^{**}_t (\sigma^{**}_t dt + dW_t),
$$
where $$\sigma^{**}_t = \lambda^*_t /\theta_t + \theta_t$$\\
 for $t\geq t_0$.

\subsection{Real-World Pricing and Benchmark-Neutral Pricing}
According to  Theorem 10.3.1 in \citeN{PlatenHe06}, one can employ the extended-market GOP $S^{**}_t$ as a num\'eraire and the real-world probability measure $P$ as the respective pricing measure to identify the minimal possible price \begin{equation}\label{Eqn3.71}
H_t = S^{**}_t E^P\bigg(  \frac{H_T}{S^{**}_T} \bigg| \mathcal{F}_t \bigg),
\end{equation} of an $\mathcal{F}_T$-measurable replicable nonnegative contingent claim $H_T$ with\\ $E^P(  \frac{H_T}{S^{**}_T} | \mathcal{F}_t )<\infty$ for $t\in[t_0,T]$. Here $E^P(  . | \mathcal{F}_t )$ denotes the conditional expectation under the real-world probability measure $P$. The above pricing formula \eqref{Eqn3.71} is known as the {\em real-world pricing formula}. \\

By minimizing the joint information of the risk-neutral pricing measure and the real-world probabiity measure, it has been shown in \citeN{Platen25b} that the so-called information-minimizing dynamics of the stock GOP $S_t$ are in the respective activity time $\tau_t$ those of a  squared radial Ornstein-Uhlenbeck process which is a genralization of the Cox-Ingersol-Ross process; see \citeN{RevuzYo99} and \citeN{CoxInRo85}. Furthermore, it has been shown in \citeN{Platen25a} that the  stock GOP when denominated in the extended-market GOP, which is the process $(S_t/S^{**}_t)_{t\ge t_0}$, forms in an information-minimizing  market a  $(P,(\mathcal{F}_t)_{t\ge t_0})$-martingale. \\ 

According to Equation 3.7 in \citeN{Platen25a},  the martingale property of \\$(S_t/S^{**}_t)_{t\ge t_0}$ allows  performing a num\'eraire change, where the stock GOP $S_t$ becomes the num\'eraire and a respective equivalent probability measure $Q_S$ the pricing measure. This gives rise to the {\em benchmark-neutral pricing formula} 
\begin{equation}\label{Eqn3.7}
H_t  = S_t E^{Q_{S}}\bigg(  \frac{H_T}{S^{}_T} \bigg| \mathcal{F}_t \bigg),
\end{equation}
where $Q_{S}$ is called the {\em benchmark-neutral  pricing measure}. Here $ E^{Q_{S}}(  . | \mathcal{F}_t )$ denotes the conditional expectation under $Q_S$, and the stock GOP $S_t$ serves as the num\'eraire for $t_0\leq t\leq T<\infty$.\\

As shown in \citeN{Platen25a}, the process $\bar W_t$, satisfying the stochastic differential $$d\bar W_t = dW_t + \frac{\lambda^*_t}{\theta_t}dt,$$ forms a  $(Q_{S},\underline{\cal{F}})$-Brownian motion. Under  $Q_S$, the stock-GOP volatility \begin{equation}\label{theta2'''}
\theta_{t}=\sqrt{\frac{4e^{ \tau_t}a_t}{S_t}}
\end{equation} is that of a squared Bessel process of dimension four in the activity time $$\tau_t=\tau_{t_0}+\int_{t_0}^{t}a_sds$$  so that \eqref{Eqn2.1} becomes
\begin{equation}
\label{Eqn3.9}
dS_t = S_t\theta_t (\theta_t dt + d\bar W_t)=4e^{\tau_t} a_t  dt + \sqrt{S_t4e^{\tau_t} a_t } d\bar W_t
\end{equation}
for $t\geq t_0$. By application of the It\^o formula it follows
$$
\tau_t = \ln \left( [\sqrt{S_.}]_{t} + \tau_{t_0} \right),
$$
which confirms Equation \eqref{tau} for the case when the step size of the time discretization for the observation of the stock GOP converges to zero.\\
The transition probability density of a squared Bessel process is that of a non-central chi-square distribution. Consequently,  the value $S_t$ of the stock GOP has in activity time under the benchmark-neutral pricing measure $Q_S$ a non-central chi-square distribution; see \citeN{Platen25a}.   For the contingent claim $H_T=1$, which pays at maturity $T$ one unit of the savings account, this leads by the benchmark-neutral pricing formula \eqref{Eqn3.7} to the  ZCB price
\begin{equation}\label{Eqn3.8}
\bar P(t,T)  = S_t E^{Q_{S}}\bigg(  \frac{1}{S^{}_T} \bigg| \mathcal{F}_t \bigg)=1-\exp\left\{-\frac{S_t}{2(e^{\tau_T}-e^{\tau_t})}\right\}
\end{equation}
for $t\in[t_0,T)$; see Equation 13.3.5 in \citeN{PlatenHe06}. For the respective hedging portfolio, the fraction $\bar\pi_t$ invested at time $t\in[t_0,T)$ in the stock GOP amounts to
$$\bar\pi_t=\frac{\frac{\partial \bar P(t,T)}{\partial S_t}S_t}{\bar P(t,T)}=\left(1-\frac{1}{\bar P(t,T)}\right)\ln(1-\bar P(t,T)).$$
Since the activity time exhibits some randomness but evolves on average linearly, the paper  employs for the generation of potential FLVRs the AZCB  payoff \eqref{HT}. It approximates the AZCB price by \eqref{barP} and the AZCB fractions by \eqref{pi}.  One notes that  these quantities approximate those of the ZCB, which is approximated by the AZCB when the  step size of the time discretization converges to zero. The extreme accuracy of the hedge portfolios in the above empirical study 
 corroborates the theory given in \citeN{Platen25b}, namely, that the predicted information-minimizing  stock GOP dynamics are close to the  prevailing dynamics of the  S\&P500 when employed as proxy of the stock GOP. \\

\section*{Conclusion}
The paper demonstrates that some classical arbitrage opportunities, like free lunches with vanishing risks, are likely to exist in the real stock market. These findings question the validity of the key assumption of the Fundamental Theorem of Asset Pricing, which provides the theoretical underpinning of the classical  mathematical finance theory. The benchmark approach offers a more general mathematical framework, where the existence of free lunches with vanishing risk does not constitute a theoretical or practical problem.

 \section*{Data}
The files used to perform the study and generate the figures of the paper is available at Harvard Dataverse https://doi.org/10.7910/DVN/YK7SX9.
\section*{Disclosure of Interest}
There are no interests to declare. No funding was received.

\bibliographystyle{chicago}
\bibliography{my} 

\begin{thebibliography}{}

\bibitem[\protect\citeauthoryear{Barone-Adesi, Platen, and Sala}{Barone-Adesi
  et~al.}{2024}]{BaroneadesiPlSa24}
Barone-Adesi, G., E.~Platen, and C.~Sala (2024).
\newblock Managing the shortfall risk of target date funds by overfunding.
\newblock {\em Journal of Pension Economics and Finance\/}, 1--25.

\bibitem[\protect\citeauthoryear{{Bloomberg Financial Services}}{{Bloomberg
  Financial Services}}{2025}]{SP500}
{Bloomberg Financial Services} (2025).
\newblock {S\&P500, Total Return Indices, Gross of Dividends}.
\newblock {Retrieved March 12, 2025}.

\bibitem[\protect\citeauthoryear{{Board of Governors of the Federal Reserve
  System (US)}}{{Board of Governors of the Federal Reserve System
  (US)}}{2025}]{DTB3}
{Board of Governors of the Federal Reserve System (US)} (2025).
\newblock {3-Month Treasury Bill Secondary Market Rate, Discount Basis [DTB3]}.
\newblock {https://fred.stlouisfed.org/series/DTB3. Retrieved March 12, 2025
  from FRED, Federal Reserve Bank of St. Louis.}

\bibitem[\protect\citeauthoryear{Cox, Ingersoll, and Ross}{Cox
  et~al.}{1985}]{CoxInRo85}
Cox, J.~C., J.~E. Ingersoll, and S.~A. Ross (1985).
\newblock A theory of the term structure of interest rates.
\newblock ~{\em 53}, 385--407.

\bibitem[\protect\citeauthoryear{Delbaen and Schachermayer}{Delbaen and
  Schachermayer}{1998}]{DelbaenSc98}
Delbaen, F. and W.~Schachermayer (1998).
\newblock The fundamental theorem of asset pricing for unbounded stochastic
  processes.
\newblock {\em Math. Ann.\/}~{\em 312}, 215--250.

\bibitem[\protect\citeauthoryear{Fergusson and Platen}{Fergusson and
  Platen}{2023}]{FergussonPl23}
Fergusson, K. and E.~Platen (2023).
\newblock Less-expensive valuation of long-term annuities linked to mortality,
  cash and equity.
\newblock {\em Annals of Actuarial Science\/}~{\em 17\/}(1), 170--207.

\bibitem[\protect\citeauthoryear{Filipovi\'c and Platen}{Filipovi\'c and
  Platen}{2009}]{FilipovicPl09}
Filipovi\'c, D. and E.~Platen (2009).
\newblock Consistent market extensions under the benchmark approach.
\newblock {\em Mathematical Finance\/}~{\em 19\/}(1), 41--52.

\bibitem[\protect\citeauthoryear{Hogg, McKean, and Craig}{Hogg
  et~al.}{2019}]{Hogg2019}
Hogg, R.~V., J.~W. McKean, and A.~T. Craig (2019).
\newblock {\em Introduction to Mathematical Statistics}.
\newblock Pearson Education Inc.,Boston, 8th edition.

\bibitem[\protect\citeauthoryear{{IFRS Foundation}}{{IFRS
  Foundation}}{2022}]{IFRS17}
{IFRS Foundation} (2022).
\newblock {IFRS 17 Insurance Contracts}.

\bibitem[\protect\citeauthoryear{Jarrow}{Jarrow}{2022}]{Jarrow22}
Jarrow, R. (2022).
\newblock {\em Continuous-Time Asset Pricing Theory}.
\newblock Springer-Finance, Second Edition.

\bibitem[\protect\citeauthoryear{Karatzas and Shreve}{Karatzas and
  Shreve}{1998}]{KaratzasSh98}
Karatzas, I. and S.~E. Shreve (1998).
\newblock {\em Methods of Mathematical Finance}.
\newblock Springer.

\bibitem[\protect\citeauthoryear{Kloeden and Platen}{Kloeden and
  Platen}{1999}]{KloedenPl92}
Kloeden, P.~E. and E.~Platen (1999).
\newblock {\em Numerical Solution of Stochastic Differential Equations}.
\newblock Springer.

\bibitem[\protect\citeauthoryear{Platen}{Platen}{2024}]{Platen25a}
Platen, E. (2024).
\newblock Benchmark-neutral pricing.
\newblock Technical report.
\newblock https://ssrn.com/abstract=4786090; arXiv:2407.01542.

\bibitem[\protect\citeauthoryear{Platen}{Platen}{2025}]{Platen25b}
Platen, E. (2025).
\newblock Information-minimizing stationary financial market dynamics.
\newblock Technical report.
\newblock https://arXiv: 2507.18395.

\bibitem[\protect\citeauthoryear{Platen and Heath}{Platen and
  Heath}{2006}]{PlatenHe06}
Platen, E. and D.~Heath (2006).
\newblock {\em A Benchmark Approach to Quantitative Finance}.
\newblock Springer.

\bibitem[\protect\citeauthoryear{Popper}{Popper}{2002}]{Popper02}
Popper, K.~R. (2002).
\newblock {\em The Logic of Scientific Discovery}.
\newblock Routledge.
\newblock Originally published as \textit{Logik der Forschung}, Vienna, 1935.

\bibitem[\protect\citeauthoryear{Revuz and Yor}{Revuz and
  Yor}{1999}]{RevuzYo99}
Revuz, D. and M.~Yor (1999).
\newblock {\em Continuous Martingales and Brownian Motion\/} (3rd ed.).
\newblock Springer.

\end{thebibliography}
\end{document}